\documentclass[preprint,aps, pra,showkeys,showpacs,preprintnumbers,amsmath,amssymb]{revtex4-1}
\usepackage[OT4]{fontenc} 
\usepackage{graphicx}
\usepackage{dcolumn}
\usepackage{bm}

\begin{document}

\preprint{APS/123-QED}

\title{Statistical properties of one dimensional Bose gas}

\author{Przemys\l aw Bienias}
\affiliation{
Center for Theoretical Physics, Polish Academy of Sciences, \\
Al. Lotnik\'ow 32/46, 02-668 Warsaw, Poland
}

\author{Krzysztof Paw\l owski}
\affiliation{%
Center for Theoretical Physics, Polish Academy of Sciences, \\
Al. Lotnik\'ow 32/46, 02-668 Warsaw, Poland
}

\author{Mariusz Gajda}
\affiliation{%
Institute of Physics, Polish Academy of Siences, \\
Al. Lotnik\'ow 32/46, 02-668 Warsaw, Poland
}

\author{Kazimierz Rz\k{a}\.{z}ewski}
\affiliation{%
Center for Theoretical Physics, Polish Academy of Sciences, \\
Al. Lotnik\'ow 32/46, 02-668 Warsaw, Poland
}
\affiliation{Faculty of Mathematics and Sciences, Cardinal Stefan Wyszy\'nski University, \\
ul. Dewajtis 5, 01-815, Warsaw, Poland}%

\date{\today}

\begin{abstract}
Monte Carlo method within, so called, classical fields approximation is applied to one dimensional weakly interacting repulsive Bose gas trapped in a harmonic potential. Equilibrium statistical properties of the condensate are calculated within a canonical ensemble. We also calculate experimentally relevant low order correlation functions of the whole gas.
\end{abstract}

\pacs{67.85.Bc, 67.85.-d, 03.75.Hh, 05.30.Jp}
                             
\keywords{ultracold atoms, statistics}
                 
\maketitle

\section{\label{sec:intro}Introduction}

Statistical properties of multiparticle quantum systems are at the heart of statistical physics. The best developed are both experiments and the theory of statistical properties of photons. In quickly developing physics of quantum degenerate dilute gases there are two aspects that make the problem of thermal equilibrium atom statistics notably different than that of photons. First, atoms collide and their interaction is necessary for the thermalization process - their approach to thermal equilibrium. Second, total number of atoms is strictly conserved and the experiments on Bose-Einstein condensation are performed with almost perfect isolation of the system from an exchange of particles with the outside world. Moreover, the system often consists of as few as a couple of hundred atoms, thus it is far from a thermodynamic limit.
Until now, fully understood is the statistics of an ideal Bose gas only \cite{navez1997,schrodinger}. In a series of papers it was found that the three commonly used statistical ensembles (grand canonical, canonical and microcanonical) while giving identical prediction for the extinction of the condensate with increasing temperature, differ in the predicted fluctuations of the number of condensed atoms. In particular these fluctuations calculated via grand canonical ensemble are absurdly large. It is worth noticing that this last observation was already made by E. Schr\"odinger \cite{schrodinger}. 
The statistics of a weakly interacting Bose gas is still a challenge. Nearly all papers deal with an academic problem of a system of N atoms confined in a rectangular box with periodic boundary conditions. There are two important simplifying aspects of such a confinement. First, even in the presence of (repulsive) interaction, the condensate wave function is still the zero momentum component of the atomic field just as for the ideal gas. Second is a simplicity of the Bogoliubov quasiparticle excitations spectrum and an analytic relation between the annihilation and creation operators of quasiparticles and the corresponding operators for atoms \cite{bogoliubov1947}. Thus, in the Bogoliubov approximation, the total Hamiltonian may be written as: 
\begin{equation}
H_B = E_0 + \sum _{\bm{k}} \epsilon(\bm{k}) b_{\bm{k}}^{\dagger } b_{\bm{k}}
\label{eqn:energyBog}
\end{equation}
or a sum of the energy of the condensate $E_0$ and the sum of the independent energies of quasiparticles (created and annihilated by $b_{\bm{k}}^{\dagger }$ and $b_{\bm{k}}$ respectively) which in this approximation are treated as independent bosons with simple and well understood equilibrium thermodynamic properties. Thus ignoring the changing number of background condensed atoms one can compute the statistics of the excited thermal atoms just from the statistics of the independent quasiparticles \cite{giorgini1998}. This way one gets a reliable results for low temperatures only. At higher temperatures the Bogoliubov spectrum gets modified and it takes a Bogoliubov-Popov form \cite{popov1983}. In this approximation the spectrum depends on the number of condensed atoms rather than on their total number. Moreover, also $E_0$ depends on $N_0$. So both parameters depend on the random variable. This presents a serious technical difficulty \cite{svidzinsky2006}. It has been overcome in a self-consistent way in \cite{idziaszek2009}. But none of these papers takes fully into account the higher order terms in the interaction Hamiltonian. Physically these terms describe the interaction between quasiparticles that leads to their instability. 
In a recent paper \cite{witkowska2010} we have shown how to calculate the statistical properties of the weakly interacting Bose gas retaining full value of the interaction energy. To this end we have proposed to use the so called classical fields approximation \cite{kagan1997}. In this approximation all long wave-length degrees of freedom of the atomic field are stripped-off their operator character and are described by complex amplitudes. The question if practically all atoms may be accounted for within the classical fields has been answered in affirmative. In a recent paper \cite{witkowska2009} we have shown that with a proper choice of the short wave-length cut-off all statistical properties of an ideal Bose gas may be reproduced using the classical fields. In \cite{witkowska2010} we have applied the classical fields to describe statistical properties of a weakly interacting bosons again confined in a box with the periodic boundary conditions.
In this paper we extend the Monte Carlo method to an experimentally relevant one dimensional weakly interacting Bose gas trapped in a harmonic potential. The results are obtained with the help of the classical fields, but are accounting fully for the interaction energy. In this case there are both theoretical \cite{petrov2000} and  experimental results \cite{dettmer2001} for the limited phase coherence of a very cold atomic sample, known as a quasicondensate and also recent measurements of the local density fluctuations \cite{esteve2006}.
In Section \ref{sec:method} we formulate the problem and describe the proposed scheme based on the classic Metropolis algorithm. 
In Section \ref{sec:stats} we present statistical properties of the condensate.
In Section \ref{sec:corr} we present the results for low order correlation functions of the whole gas. 

\section{\label{sec:method}The method}
We study a one dimensional, repulsive, weakly interacting Bose gas confined in a harmonic trap. Excellent realizations of such a system are available \cite{esteve2006}
Thus, our Hamiltonian of the one dimensional Bose gas has a form:
\begin{eqnarray}
\nonumber H&=& \int \hat{\Psi}^{\dagger }(x)\left( \frac{p^2}{2m}+\frac{1}{2}m\,\omega^2 x^2  \right) \hat{\Psi}(x) dx +\\
& & +\,\frac{g}{2} \int \hat{\Psi}^{\dagger }(x) \hat{\Psi}^{\dagger }(x) \hat{\Psi}(x) \hat{\Psi}(x) ,
\label{eqn:hamiltonian}
\end{eqnarray}
The Hamiltonian is a sum of the single particle oscillator energy with mass $m$ and angular frequency $\omega$ and a conventional contact interaction with the coupling constant $g$. 
A convenient base is provided by the eigenstates of the harmonic oscillators $\phi_n(x)$:
\begin{equation}
\hat{\Psi}(x) = \sum _{n} \phi_n(x) \hat{a}_n 
\label{eqn:ffalowa}
\end{equation}
where $\hat{a}_n$ annihilates an atom in the $n$-th harmonic oscillator state. The classical field approximation consists in replacing the creation and annihilation operators by classical complex amplitudes:
\begin{equation}
\hat{a}_n, \hat{a}_n^{\dagger} \mapsto \alpha_n, \alpha_n ^{*}
\label{eqn:mapowanie}
\end{equation}
In \cite{witkowska2009} we have shown, that probabilistic properties of the condensate for one dimensional ideal Bose gas in a harmonic trap are perfectly reproduced by the classical fields approximation provided the number of degrees of freedom is kept finite with the last retained state chosen as:
\begin{equation}
K = \frac{k_B T}{\hbar \omega}
\label{eqn:kvst}
\end{equation}
where $T$ is the absolute temperature and $k_B$ is the Boltzmann constant. This suggests the classical fields approach may be used also for the weakly interacting gas \cite{witkowska2010}. In this approximation the energy of a given configuration of the field is given as:
\begin{equation}
E\left( \left\{ \alpha_i \right\} \right) = \hbar \omega \sum_{n=0}^K n |\alpha_n |^2 + E_{int}\left( \left\{ \alpha_i \right\} \right)
\label{eqn:energy}
\end{equation}

where $E_{int}\left( \left\{ \alpha_i \right\} \right)$ is the quartic polynomial in the amplitudes $\alpha$.
Hence, the probability distribution of a given configuration of the classical field according to the canonical ensemble is:
\begin{equation}
P\left( \left\{ \alpha_i \right\} \right) = \frac{1}{Z(N,T)} \exp \left[-\frac{E\left( \left\{ \alpha_i \right\} \right)}{k_B T}\right]
\label{eqn:probability}
\end{equation}
where $Z(N,T)$ is the classical partition function for $N$ atoms at temperature $T$. We need to generate this probability distribution with a set of the amplitudes subject to the constraint given by the fixed number of particles:
\begin{equation}
\sum^K_{n=0}|\alpha_n|^2=N
\label{eqn:normalization}
\end{equation}
The best known Monte Carlo realization of this distribution is given by the Metropolis algorithm \cite{metropolis1953}. Before we turn to the results there are two important remarks: For interacting atoms the condensate wave function is no longer the empty harmonic potential ground state, thus its occupation is not $|\alpha_0|^2$. Instead, following \cite{penrose1956},  the identification of the condensate requires a diagonalization of the single particle density matrix, which in our harmonic oscillator representation is:
\begin{equation}
\rho_{i,j}=\langle\alpha_i^* \alpha_j\rangle=\sum_n\lambda_n\, \beta_i^*(n) \beta_j(n)
\label{eqn:rho}
\end{equation}
where the mean value is taken with the probability distribution \eqref{eqn:probability} and the eigenvector corresponding to the dominant eigenvalue is the condensate.
The other remark concerns the cut-off condition \eqref{eqn:kvst}. For the interacting gas we need a higher cut-off. We know that the repulsive gas at zero temperature is broader then the ground state of the harmonic potential thus its wave function needs higher energy terms. Remembering that in the Bogolubov approximation the quasiparticle excitations have energies counted not from zero but from the chemical potential $\mu$ we postulate the modification of the cut-off condition in the form:
\begin{equation}
K\hbar \omega=\mu +k_BT
\label{eqn:cutoff}
\end{equation}
which will be independently verified in the next Section.

\section{\label{sec:stats} Statistical properties of 1D condensate}
Equipped with the numerical scheme described in some detail in the preceding Section we turn now to the discussion of results. We begin with statistical properties of the condensate. 
The only earlier results were obtained with the help of the time dependent methods \cite{kadio2005, proukakis2006}. In this paper we present the first results obtained with the methods of equilibrium statistical mechanics.
In our classical fields approximation, however, effects of quantum fluctuations, such as quantum depletion are missing. Throughout this paper we use the oscillator units of position, energy and temperature: $\sqrt{\frac{\hbar}{m\omega}}$, $\hbar \omega$, $\frac{\hbar\omega}{k_B}$ respectively. All calculations are performed for $500$ atoms. Our main parameter is just the coupling constant $g$ -- all simulations are done for $g=0.02$ (what corresponds to the $^{87}$Rb in our units) and $g=1$. As a reference point we always have the ideal gas case. For the ideal gas we have an exact probability distribution of the number of uncondensed atoms $N_{ex}$ (thus with remaining $N_0 = N-N_{ex}$ in the condensate):
\begin{equation}
 P\left(N_{ex}, T\right) = \frac{1}{\xi^{N_{ex}}}\prod_{l=N_{ex}+1}^N \left(1-\xi^l\right),
\end{equation}
where  $\xi=\exp\left(-\hbar\omega/k_B T\right)$.
We also have its best classical fields approximation:
\begin{equation}
 P_{cl}\left(N_{ex}, T\right) = \frac{1}{1-\xi^{N}}\left( \frac{ 1-\xi^{N_{ex}}}{1-\xi^N} \right) ^{K-1}
\label{eqn:excited}
\end{equation}
with the cut-off parameter $K$ chosen according to \eqref{eqn:kvst} and the Monte Carlo representation of the classical distribution. The detailed derivation of \eqref{eqn:excited} is given in the Appendix. 

From these, coinciding reference points we are then departing to the largely unchartered territory of the interacting gas.
\begin{figure}
\includegraphics[width= 8.6cm, angle=0, clip=true]{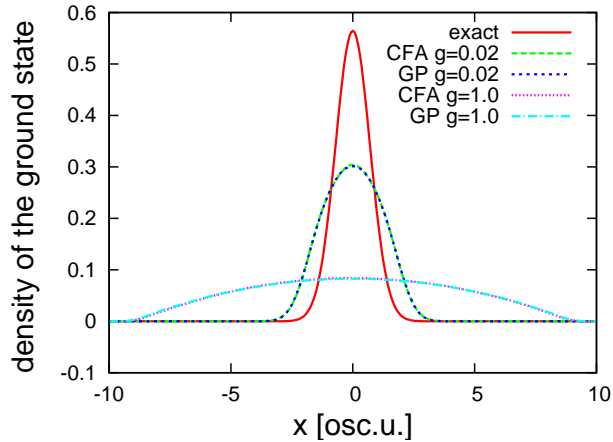}
\caption{(Color online)Zero temperature spatial distribution of the ground state for various values of the parameter $g$. Comparison between the classical field approximation (data from Metropolis algorithm) and Gross-Pitaevski equation (from imaginary time evolution). The red solid line corresponds to exact result for non-interacting gas.}
\label{fig:shape}
\end{figure}
In Fig. \ref{fig:shape} we plot the zero temperature spatial distribution of the condensate for several values of the coupling. Note the standard broadening of the condensate wave function. At this point our method merely seeks the minimum of the classical energy functional \eqref{eqn:probability} satisfying the constraint \eqref{eqn:normalization}. In fact this is nearly the same as the ground state of the Gross-Pitaevskii equation conveniently computed by the imaginary time propagation.
\begin{figure}
\includegraphics[width= 8.6cm, angle=0, clip=true]{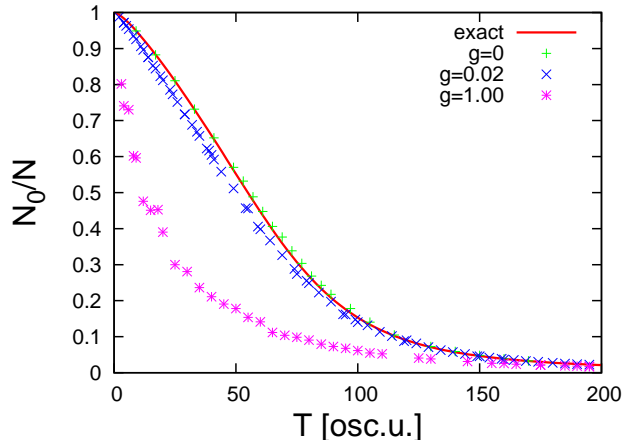}
\caption{(Color online)Relative average number of atoms in the ground state as a function of temperature $T$. The red solid line corresponds to the exact result for non-interacting gas, points -- results of classical field approximation.}
\label{fig:n0}
\end{figure}
Next we look at the mean number of condensed atoms as a function of temperature, Fig \ref{fig:n0}. In the 1D case this number decreases gradually even in $N\rightarrow \infty$ limit. As it should be, the exact ideal gas result (solid, red online) is reproduced very well by the corresponding classical field Monte Carlo results (green crosses).  Note that for a stronger coupling the depletion of the condensate with growing temperature becomes very rapid. This plot provides a direct test of our proposed modification of the cut-off condition for the interacting gas. 
The shift of the cut-off energy by the chemical potential is the smallest increase that guaranties $\frac{N_0}{N}\xrightarrow [T\rightarrow 0]{} 1$ and in the same time gives correct zero temperature condensate wave function. 
\begin{figure}
\includegraphics[width= 8.6cm, angle=0, clip=true]{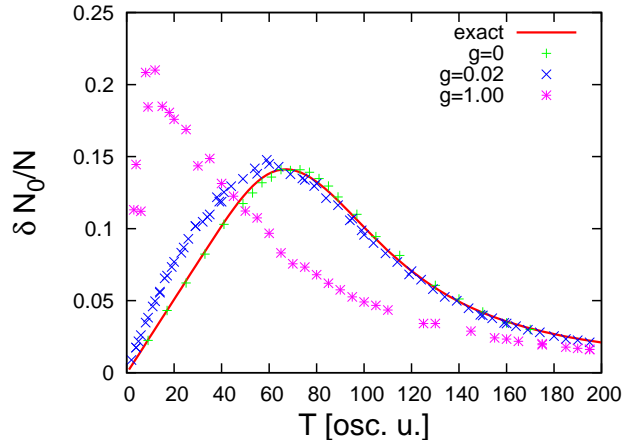}
\caption{(Color online)Relative fluctuation of number of atoms in the ground state.  The red solid line corresponds to exact result for non-interacting gas, points -- results of classical field approximation. }
\label{fig:n0fluct}
\end{figure}
Next let us look at fluctuations. It is the fluctuations that were ensemble dependent for the ideal gas. 
In Fig. \ref{fig:n0fluct} we plot the variance of the condensate occupation probability distribution for the interacting gas again compared to the ideal gas canonical result (solid, red online). We see the shift of the curve towards lower temperatures. As for all finite systems, it is not easy to define a characteristic cross-over temperature based on the depletion plots as in Fig. \ref{fig:n0}. It is natural to define such a cross-over characteristic temperature as the one corresponding to the maximum variance \cite{idziaszek2003}.
\begin{figure}
\includegraphics[width= 8.6cm, angle=0, clip=true]{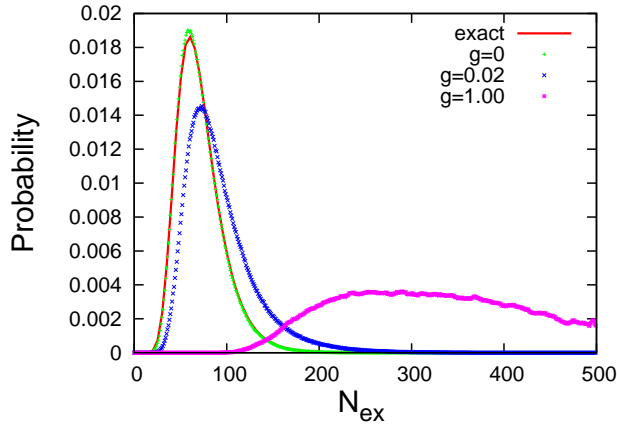}
\caption{(Color online)Probability distribution of the excited states' occupation. Solid red line -- exact results for non-interacting gas, symbols -- results of the classical field approximation.}
\label{fig:nexc}
\end{figure}
Of course our Monte Carlo method gives also access to the full probability distribution of the occupation of the condensate state. For $T=20$ this is shown in Fig. \ref{fig:nexc}. 

\section{\label{sec:corr}Low order correlation functions of 1D Bose gas}

In spite of all the theoretical effort, until now there are no experimental measurements of the temperature dependent statistical properties of the condensate. The relative fluctuations are significant only for small samples of several hundred atoms. Sufficiently precise determination of the condensate fraction with nearly perfect control of the total number of trapped atoms remains a challenge.
However, measured indirectly were temperature dependent coherence properties of nearly 1D Bose gas \cite{kreuzmann2003} and also local density fluctuations of such a gas \cite{esteve2006}. Those are of course, also accessible to us.
There are two remarks in order: First, our cut-off was optimized to reproduce statistics of the condensate. This way it is not optimized for the thermal atoms. In fact, as was shown in \cite{witkowska2009}, it overestimates the population of thermal modes. 
It must be so since in the classical field approximation all atoms are distributed over a finite number of low-lying states instead of the infinite number of them. Thus, particularly at high temperatures, when most atoms are thermal, we expect a deteriorating accuracy of the method.
Second is the problem of ordering. Of course within the classical fields approximation  fields do commute. Thus, there is no difference between the density-density correlation function and the fourth order normally ordered correlation function:
\begin{displaymath}
 \langle \hat{\Psi}^{\dagger}\hat{\Psi}\,\hat{\Psi}^{\dagger} \hat{\Psi}\rangle = \langle \hat{\Psi}^{\dagger}\hat{\Psi}^{\dagger} \hat{\Psi}\hat{\Psi}\rangle + \langle \hat{\Psi}^{\dagger} \hat{\Psi}\rangle.
\end{displaymath}
But at very high temperatures the last term, missing for classical fields, becomes dominant. It gives rise to what experimenters call the shot noise. Classical fields cannot possibly reproduce this contribution unless it would be introduced by hand.
The most interesting property of the lowest order correlation function for the 1D Bose gas has been first noted in \cite{petrov2000}. Unlike in 3D, the coherence length in 1D gas may be shorter than the length of the condensate. This phenomenon is called a quasicondensate. In the experiment \cite{dettmer2001} it has been observed indirectly: Upon expansion, quasicondensate's phase fluctuations turn into density fluctuations that can be directly observed in standard absorptive imaging.
\begin{figure}
\includegraphics[width= 8.6cm, angle=0, clip=true]{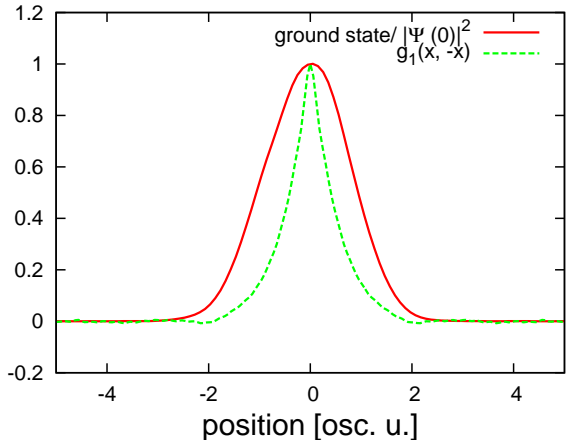}
\caption{(Color online)The first order correlation function and the density of the ground state. For easier comparison the quasicondensate density profile has been rescaled by the values at the center of the trap.}
\label{fig:g1shape}
\end{figure}
In Fig. \ref{fig:g1shape} we show the comparison of the condensate wave-function and the lowest order correlation function
\begin{displaymath}
g_1 (-x, x) = \frac{\langle\Psi^* (-x) \Psi (x) \rangle }{\langle\left|\Psi(x) \right|^2\rangle}
\end{displaymath}
in the quasicondensate regime. 
\begin{figure}
\includegraphics[width= 8.6cm, angle=0, clip=true]{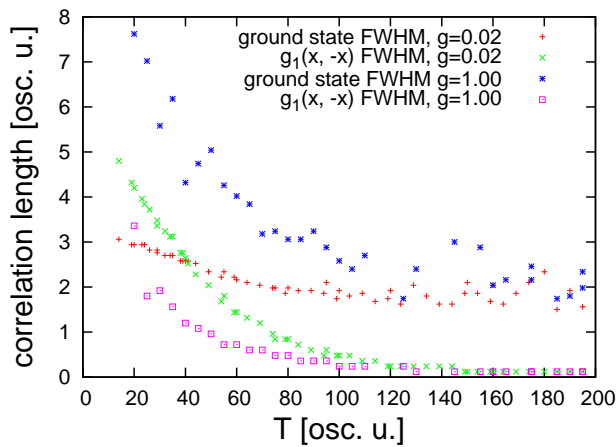}
\caption{(Color online)The correlation length and the size of the ground state wave function for two different values of interaction. Both lengths defined as full width at half maximum.}
\label{fig:corl}
\end{figure}
More quantitative analysis is shown in Fig. \ref{fig:corl}. Here we vary the temperature confronting the length of the condensate with the correlation length for our two standard coupling constants. Note the crossing of the two curves defining the onset of the quasicondensate. We add that the 1D ideal gas has no quasicondensation. 
\begin{figure}
\includegraphics[width= 8.6cm, angle=0, clip=true]{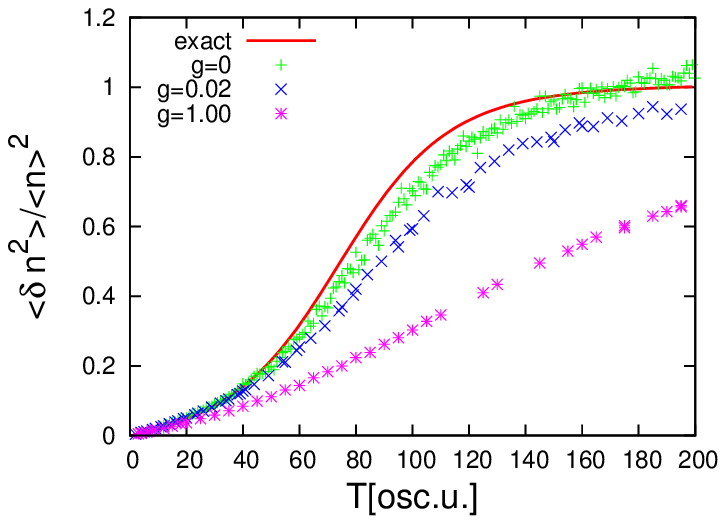}
\caption{(Color online)Normalized fluctuations of the atoms' number at the center of the trap as a function of the temperature. The red solid line corresponds to exact result for non-interacting gas, points -- results of classical field approximation. Note imperfect agreement of the classical fields for an ideal gas.}
\label{fig:fluct0}
\end{figure}

Finally, we present results for the local density fluctuations. They were measured recently in Hanover \cite{kreuzmann2003} and in Orsay \cite{esteve2006}. In Fig. \ref{fig:fluct0} we present temperature dependence of the local density fluctuations at the center of the trap defined as
\begin{displaymath}
\frac{\langle\delta^2 n\rangle}{\langle n\rangle^2} = \frac{\langle \left|\Psi (0)\right| ^4 \rangle - \langle \left|\Psi (0)\right| ^2 \rangle}{\langle \left|\Psi (0)\right| ^2 \rangle^2}
\end{displaymath}
Unlike in the statistical properties of the condensate itself, even for the ideal gas there is certain difference between the exact (solid, red online) and the Monte Carlo classical fields. As mentioned above it is a result of the distorted occupation of the thermal modes in the classical fields approach. The interaction reduces the density fluctuations as predicted in \cite{kheruntsyan2003}.
\begin{figure}
\includegraphics[width= 8.6cm, angle=0, clip=true]{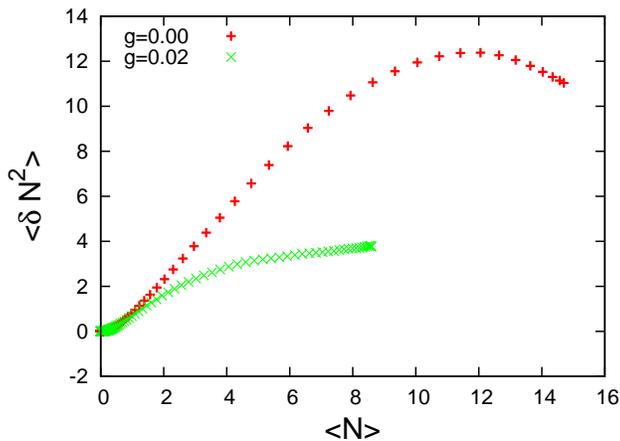}
\caption{(Color online)Local density fluctuations versus local density at the temperature $T=20$. As the strength of interaction increases, the bosonic gas enters the quasicondensate regime.}
\label{fig:fluctvsDens}
\end{figure}
In Ref. \cite{esteve2006} the density fluctuations were measured \textit{in situ} across a nearly one-dimensional sample versus the local density. The comparison of the ideal gas and the interacting gas of the same dependence is presented in Fig. \ref{fig:fluctvsDens}. It has the main qualitative properties just like in the experiment again showing the reduction of the fluctuations due to the interactions. The quantitative comparison is hard since experimental parameters of the number of particles or the temperature are known with large error bars.
The authors of \cite{esteve2006} explored also a very high temperature regime with the density fluctuations dominated by the shot noise. This is a regime not accessible to classical fields for two related reasons: in this regime all states have a very small, fractional occupation and also classical fields are missing the quadratic term resulting from the commutator of the Bose fields, term essential in the high temperature regime.
\section{\label{sec:concl} Conclusions}
In this paper we have demonstrated a power of the novel computational tool to tackle the equilibrium thermodynamics of weakly interacting gas consisting of Bose particles. The method, based on classical fields approximation to the full quantum theory exploits classical Monte Carlo technique to explore the phase space of the underlying finite dimensional classical system, for which the set of canonical variables replaces the annihilation and creation operators. 

We have applied the method to the 1D repulsive gas trapped in a harmonic potential calculating not only the statistical properties of the condensate but also the low-order correlation functions of the whole gas that may be directly confronted with the measurements.

\begin{acknowledgments}
The authors appreciate discussions with Emilia Witkowska. We also acknowledge financial support of Polish Government Funds for 2010-2012.
\end{acknowledgments}

\appendix*
\section{\label{appendix}Partition function for the ideal Bose gas in a harmonic trap -- classical fields}

The quantum partition function in the canonical ensemble may be written as follows:
\begin{equation}
 Z\left(N, \beta\right) = \sum_{n_0=0}^{\infty} \sum_{n_1=0}^{\infty} \ldots \sum_{n_k=0}^{\infty}\ldots e^{-\beta \hbar \omega \sum_{k=0}^{\infty}k\,n_k} \delta_{\sum_{k}n_k\,, N} ,
\label{aeqn:qpart}
\end{equation}
where $\beta = 1/k_B T$ and $\omega$ is a frequency of the harmonic trap. The Kronecker delta in Eq. \ref{aeqn:qpart} enforces in the conservation of the total number of particles. In the classical version the sums have to be replaced with integrals $\sum_{n_k=0}^{\infty}\mapsto \frac{1}{\pi}\int \mbox{d}^2\alpha_k $ and the Kronecker's delta turns into the Dirac's delta $\delta_{\sum_{k=0}^{\infty}n_k\,, N} \mapsto \delta\left( \sum_{n=0}^{\infty} \left| \alpha_n\right|^2 - N\right)$. Thus classical counterpart of the above formula reads:
\begin{widetext}
\begin{eqnarray}
 \nonumber Z\left(N, \beta\right) &=& \int \frac{\mbox{d}^2\alpha_0}{\pi}\ldots \int \frac{\mbox{d}^2\alpha_K}{\pi} e^{-\beta \hbar \omega \sum_{k=0}^{K}k\left| \alpha_k\right|^2 }\delta\left( \sum_{n=0}^{K} \left| \alpha_n\right|^2 - N\right) =\\
 &=& \prod_{j=0}^K \int \frac{\mbox{d}^2\alpha_j}{\pi}\,e^{-\beta \hbar \omega j\left| \alpha_j\right|^2} \delta\left( \sum_{n=0}^{K} \left| \alpha_n\right|^2 - N\right),
\end{eqnarray}
\end{widetext}
where we have also taken into account the cut-off $K$, defined in \eqref{eqn:kvst}. When we use the integral form of Dirac's delta $\delta (x) = \frac{1}{2\pi}\int_{-\infty}^{\infty}\mbox{d}\eta \,\exp\left(-\dot{\imath}\eta x\right)$, the classical partition function takes the form
\begin{equation}
  Z\left(N, \beta\right) = \frac{1}{2\pi}\int_{-\infty}^{\infty}\mbox{d}\eta\,e^{-\dot{\imath}N\eta}\prod_{j=0}^K \int \frac{\mbox{d}^2\alpha_j}{\pi}\,e^{-\left(\beta \hbar \omega j-\dot{\imath}\eta\right)\left| \alpha_j\right|^2} .
\label{aeqn:claspart}
\end{equation}

In the above we can easily compute all integrals over amplitudes $\alpha_k$
\begin{equation}
  Z\left(N, \beta\right) = \frac{1}{2\pi}\int_{-\infty}^{\infty}\mbox{d}\eta\,e^{-\dot{\imath}N\eta}\prod_{j=0}^K \frac{1}{\beta \hbar \omega j-\dot{\imath}\eta }, 
\label{aeqn:help1}
\end{equation}
and then the appropriate contour integral over $\eta$  which has $K+1$ poles:
\begin{equation}
  Z\left(N, \beta\right) = \frac{1}{\left(\beta\hbar\omega\right)^K}\sum_{j=0}^{K} \xi^{jN}\prod_{k\neq j}^K \frac{1}{j-k}, 
\label{aeqn:help2}
\end{equation}
where $\xi = e^{-\beta\hbar\omega}$.

Note that the expression $\prod_{k\neq j}^K \frac{1}{j-k}$ is just equal to $\frac{(-1)^j}{K!}\binom{K}{j}$, and the whole sums of products can be rewritten in the simpler form:
\begin{equation}
  Z\left(N, \beta\right) = \frac{1}{K!\left(\beta\hbar\omega\right)^K}\sum_{j=0}^{K} \left(-\xi^N\right)^j \binom{K}{j}.
\label{aeqn:help3}
\end{equation}
Using the Newton formula we can further simplify the partition function, getting finally:
\begin{equation}
  Z\left(N, \beta\right) = \frac{1}{K!}\left( \frac{1-\xi^N}{\beta\hbar\omega} \right)^K .
\label{aeqn:help4}
\end{equation}

In the same manner we can compute the partition function for excited atoms:  
\begin{displaymath}
  Z_{ex}\left(N_{ex}, T\right) = \frac{1}{(K-1)!}\left( \frac{1-\xi^{N_{ex}}}{\beta\hbar\omega} \right)^{K-1},
\end{displaymath}
where $N_{ex}$ is the number of atoms in all excited states. Thus probability of finding exactly $N_{ex}$ excited atoms in the sample of $N$ atoms at the temperature $T = K \frac{\hbar \omega}{k_B} $  is equal to
\begin{displaymath}
  P_{cl} \left(N_{ex}, T\right) =\frac{Z_{ex}\left(N_{ex}, T\right) }{Z\left(N, \beta\right)} = \frac{1}{1-\xi^{N}}\left( \frac{ 1-\xi^{N_{ex}}}{1-\xi^N} \right) ^{K-1}.
\end{displaymath}

\end{document}